\documentclass[twocolumn,prd,floatfix,amsmath,nofootinbib,amssymb,floatfix]{revtex4}
\usepackage{graphicx,color,dcolumn,booktabs,bm}
\usepackage{longtable,lscape}
\usepackage{txfonts}
\usepackage{overpic}
\usepackage{amssymb}
\usepackage{indentfirst}
\usepackage{feynmf}
\usepackage{slashed}
\usepackage{cases}
\usepackage{multirow}
\usepackage[colorlinks,
            citecolor=blue,
            anchorcolor=red,
            menucolor=red,
            linkcolor=red,
            filecolor=red,
            runcolor=red,
            urlcolor=blue,
            frenchlinks=red]{hyperref}
\usepackage{epstopdf}
\usepackage{bm}
\usepackage{makecell}
\usepackage{threeparttable}
\begin{document}
\title{Role of the newly measured $B\to KD\bar{D}$ process to establish $\chi_{c0}(2P)$ state}
\author{Ming-Xiao Duan$^{1,2}$}\email{duanmx16@lzu.edu.cn}
\author{Jun-Zhang Wang$^{1,2}$}\email{wangjzh2012@lzu.edu.cn}
\author{Yu-Shuai Li$^{1,2}$}\email{liysh20@lzu.edu.cn}
\author{Xiang Liu$^{1,2,3}$\footnote{Corresponding author}}\email{xiangliu@lzu.edu.cn}
\affiliation{
$^1$School of Physical Science and Technology, Lanzhou University, Lanzhou 730000, China\\
$^2$Research Center for Hadron and CSR Physics, Lanzhou University and Institute of Modern Physics of CAS, Lanzhou 730000, China\\
$^3$Lanzhou Center for Theoretical Physics, Key Laboratory of Theoretical Physics of Gansu Province and Frontiers Science Center for Rare Isotopes, Lanzhou University, Lanzhou 730000, China}

\begin{abstract}
In this work, the branching ratio $\mathcal{B}[B\to K\chi_{c0}(2P)]$ is extracted for the first time through the fit fractions in the newly measured $B\to KD\bar{D}$ process by LHCb. With the rescattering mechanism, this extracted branching ratio is reproduced well. Our study further enforces the role of the newly measured $B\to KD\bar{D}$ process to establish the $\chi_{c0}(2P)$ charmonium state, which is a crucial step when constructing the charmonium family.
\end{abstract}

\maketitle

\section{Introduction}\label{sec1}

In 2020, with the data of a total luminosity of 9 fb$^{-1}$ collected in $\sqrt{s}=$7, 8, and 13 TeV, the LHCb Collaboration measured the $B^+\to K^+D^+D^-$ process \cite{Aaij:2020ypa, Aaij:2020hon}. In this experimental work, through a Dalitz analysis on the $B^+\to K^+D^+D^-$ process, LHCb constructed the invariant mass spectra of the $D^+D^-$, $D^-K^+$, and $D^+K^+$ channels. In the invariant mass spectrum of $D^-K^+$, the $X_0(2900)$ and $X_1(2900)$ structures were reported \cite{Aaij:2020ypa, Aaij:2020hon}. Moreover, in the $D^+D^-$ channel, two charmonia $\chi_{c0}(3930)$ and $\chi_{c2}(3930)$ with $J^{PC}=0^{++}$ and $2^{++}$ state, respectively, were found. Their masses and widths are determined to be \cite{Aaij:2020ypa, Aaij:2020hon}
\begin{equation*}
\begin{split}
M_{\chi_{c0}(3930)}&=3.9238\pm0.0015({\rm stat})\pm0.0004({\rm syst})~{\rm GeV},\\
\Gamma_{\chi_{c0}(3930)}&=17.4\pm5.1({\rm stat})\pm0.8({\rm syst})~{\rm MeV},\\
M_{\chi_{c2}(3930)}&=3.927\pm0.0024({\rm stat})\pm0.0008({\rm syst})~{\rm GeV},\\
\Gamma_{\chi_{c2}(3930)}&=34.2\pm6.6({\rm stat})\pm1.1({\rm syst})~{\rm MeV},\\
\end{split}
\end{equation*}
which confirms the prediction from the Lanzhou group of small mass splitting and narrow width for the $\chi_{c0}(2P)$ and $\chi_{c2}(2P)$ states \cite{Duan:2020tsx}.
Especially, it is the first time to establish the $\chi_{c0}(2P)$ state in its open-charm decay channel, which also clarifies the messy situation of the $X(3915)$.

Taking this opportunity, we should briefly introduce why there exists this messy situation of the $X(3915)$. The $X(3915)$ was found by the Belle Collaboration in the $\gamma\gamma\to \omega J/\psi$ process, which has mass $m=3915\pm3({\rm stat})\pm2({\rm syst})$ MeV and width $\Gamma=17\pm10({\rm stat})\pm3({\rm syst})$ MeV \cite{Uehara:2009tx}. Later, the Lanzhou group studied the mass spectrum and decay behavior of $2P$ charmonia, and indicated that the $X(3915)$ can be assigned as a charmonium $\chi_{c0}(2P)$ \cite{Liu:2009fe}. It means that the $X(3915)$ must have $J^{PC}=0^{++}$ quantum number \cite{Liu:2009fe}, which was confirmed by the BaBar measurement \cite{Lees:2012xs}. Thus, in the 2013 version of the Particle Data Group (PDG), the $X(3915)$ was collected as the $\chi_{c0}(2P)$ state \cite{Beringer:1900zz}.

For the $\chi_{c0}(2P)$ charmonium assignment to the $X(3915)$, Guo {\it et al}. \cite{Guo:2012tv} proposed three questions: (1) why does the hidden-charm decay $X(3915)\to J/\psi \omega$ have a large width? (2) why was the dominant decay mode of $X(3915)\to D\bar{D}$ not observed in experiments? (3) why is the mass gap between the $X(3915)$ and the $Z(3930)$ far smaller than that between $\chi_{b0}(2P)$ and $\chi_{b2}(2P)$?
Based on their analysis to the invariant mass spectrum of $D\bar{D}$ in the $\gamma\gamma \to D\bar{D}$ reaction, they suggested a so-called structure around 3.8 GeV as the candidate of the $\chi_{c0}(2P)$ state \cite{Guo:2012tv}. In 2017, the Belle Collaboration measured the process $e^+e^- \to J/\psi D\bar{D}$, and claimed that a broad structure near the $D\bar{D}$ threshold named as the $X(3860)$ should be a genuine candidate of the $\chi_{c0}(2P)$ with a preferable assignment of $J^{PC}=0^{++}$ \cite{Chilikin:2017evr}.  Additionally, in Ref.~\cite{Olsen:2014maa}, some contradictions on the decay branching ratio of the $Y(3940)$ were indicated by a combined analysis for the $\gamma\gamma\to \omega J/\psi$, $\gamma\gamma\to D\bar{D}$, and $B\to K\omega J/\psi$ processes when treating the $Y(3940)$ and $X(3915)$ as the same $\chi_{c0}(2P)$ state. The above theoretical and experimental studies result in the messy situation of the $X(3915)$.

Facing to this controversial issue, the Lanzhou group still believed that the $X(3915)$ is a good candidate of the $\chi_{c0}(2P)$ state. In the past years, the Lanzhou group found solutions to three questions raised by Guo {\it et al}. \cite{Guo:2012tv} accordingly. In Ref.~\cite{Chen:2012wy}, Lanzhou group performed a combined analysis of the $D\bar{D}$ invariant mass spectrum and the angular distribution of the reaction $\gamma\gamma \to D\bar{D}$, and found that the $Z(3930)$ reported in the $D\bar{D}$ channel may contain two charmonium states with the $J^{PC} = 0^{++}$ and $2^{++}$ quantum numbers. At the same time, the calculations by considering charmed meson loops indicate that the large width of hidden-charm decay of $X(3915)\to J/\psi \omega$ can be understood well \cite{Chen:2013yxa}.
In a recent work published in 2020, Lanzhou group carried out a study under the unquenched picture to these charmonium $2P$ states \cite{Duan:2020tsx}, and indicated the coupled-channel effect is crucial to depict the charmonia $2P$ states. With the coupled-channel effect, the mass splitting between the $\chi_{c0}(2P)$ and $\chi_{c2}(2P)$ states can be decreased to be $13$ MeV, which is consistent with the mass splitting between the $X(3915)$ and the $Z(3930)$. Additionally, the predicted narrow decay width of the $\chi_{c0}(2P)$ is comparable with the decay width of the $X(3915)$ \cite{Duan:2020tsx}. In the same paper \cite{Duan:2020tsx}, the Lanzhou group also demonstrated that the charmonium
$\chi_{c0}(2P)$ must be a narrow state due to the existence of the node effect.
{Except for the results from the Lanzhou group, in the Ref.~\cite{Wang:2019evy}, the authors studied the $D\bar{D}$ mass distribution in the  $e^+e^-\to J/\psi D\bar{D}$ reaction~\cite{Chilikin:2017evr}. With the unitary formalism, they found that it was premature to claim the $X(3860)$ as a $\chi_{c0}(2P)$ state. And in their subsequent work~\cite{Wang:2020elp}, the authors showed there were no peaks around 3.86 GeV contained in the data, which are divided by the corresponding phase space in $\gamma\gamma\to D\bar{D}$ process~\cite{Uehara:2005qd, Aubert:2010ab}.} Thus, the possibility of the broad structure
$X(3860)$ reported by Belle \cite{Chilikin:2017evr} as the $\chi_{c0}(2P)$ state basically can be excluded.
As a result, all questions raised by Guo {\it et al}. \cite{Guo:2012tv} were well answered, and the conclusion of the $X(3915)$ as a $\chi_{c0}(2P)$ state is further reinforced. A key point of finally identifying the $X(3915)$ as a $\chi_{c0}(2P)$ state is to experimentally search for its $D\bar{D}$ decay mode of the $X(3915)$. The LHCb result of $B^+\to K^+D^+D^-$\cite{Aaij:2020ypa, Aaij:2020hon} mentioned above provides direct support to this point.

Obviously, it is not the end of whole study around the $X(3915)$.
The newly measured $B^+\to K^+D^+D^-$ process not only establishes the $\chi_{c0}(2P)$ state, but it also provides the information of the fit fraction of the contribution $B^+\to K^+\chi_{c0}(2P) \to K^+D^+D^-$ \cite{Aaij:2020ypa, Aaij:2020hon}, which can be applied to extract crucial information of the branching ratio $\mathcal{B}[B\to K\chi_{c0}(2P)]$. Thus, we carry out this study of extracting the branching ratio $\mathcal{B}[B\to K\chi_{c0}(2P)]$, which becomes one of the main tasks of this work.

Based on this extracted branching ratio, we may perform a deeper investigation
of the $B\to K\chi_{c0}(2P)$ decay. Towards the $B\to K\chi_{c0}(2P)$ process, since the matrix element $\langle\chi_{c0}(2P)|(c\bar{c})_{V,A}|0\rangle$=0, the amplitude vanishes under the na\"ive factorization approach, which is generally adopted in the investigation of the nonlepton $B$ decay process \cite{Bauer:1986bm, Rosner:1990xx, Luo:2001mc}. On the contrary, the extracted branching ratio of $B\to K\chi_{c0}(2P)$ is sizable. Thus, there should exist some extra mechanism to mediate the $B\to K\chi_{c0}(2P)$ process. For the similar problem, the rescattering mechanism had already been employed in discussing the $B\to K\chi_{c0}(1P)$, $B\to Kh_c(1P)$, $B\to KX(3823)$, $B\to K\eta_{c2}(1^1D_2)$, and $B\to K\psi_2(1^3D_3)$ processes~\cite{Colangelo:2002mj, Colangelo:2003sa, Xu:2016kbn}. The results show that the branching ratio of these processes can be well understood with the rescattering mechanism, which means that the rescattering mechanism cannot be ignored for the $B\to K\chi_{c0}(2P)$ decay. In this work, we try to understand the extracted branching ratio $\mathcal{B}[B\to K\chi_{c0}(2P)]$ by introducing the rescattering mechanism, which is an interesting issue around the $X(3915)$. Of course,
our study may further show the importance of the nonperturbative effect existing in the $B$ decay.

The paper is organized as follows. After the Introduction in Sec.~\ref{sec1}, we illustrate how to extract the information of the branching ratio $\mathcal{B}[B\to K\chi_{c0}(2P)]$ in Sec.~\ref{sec2}. And then, we try to understand this obtained branching ratio of $B\to K\chi_{c0}(2P)$ with the rescattering mechanism in Sec.~\ref{sec3}. The paper ends with the discussion and conclusion in Sec.~\ref{sec4}, where we further estimate the value of the product branching ratio $\mathcal{B}[B\to K\chi_{c0}(2P)]\times \mathcal{B}[\chi_{c0}(2P)\to \omega J/\psi]$, and discuss the connection of the $Y(3940)$ and $\chi_{c0}(2P)$.

\section{Extracting the branching ratio of $B\to K\chi_{c0}(2P)$ via the $B\to KD\bar{D}$ data}\label{sec2}

In the experimental analysis of the LHCb Collaboration for the $B^+\to K^+D^+D^-$ process, charmonia $\psi(3770)$, $\chi_{c0}(3930)$, $\chi_{c2}(3930)$, $\psi(4040)$, $\psi(4160)$, and $\psi(4415)$ were considered in depicting the  $D^+D^-$ invariant mass spectrum \cite{Aaij:2020ypa, Aaij:2020hon}. Through a combined fit for the invariant mass spectrum---$m(D^+D^-)$, $m(D^-K^+)$, and $m(D^+K^+)$---and the angular distribution in the $D^+D^-$ channel, LHCb not only got the resonance parameters of $\chi_{c0}(3930)$, but also determined its fit fraction. To depict the $B$ decay processes in the experiment, the differential decay width is
\begin{equation}\label{2eqDifWid}
\begin{split}
d\Gamma=\frac{1}{(2\pi)^3}\frac{1}{32m_B^3}\left|\mathcal{A}\right|^2{\rm d}m_{12}^2{\rm d}m_{23}^2,
\end{split}
\end{equation}
where $\mathcal{A}$ is the amplitude for the corresponding three-body decay process. With a coherent sum of amplitudes from resonant or nonresonant contribution, the total amplitude $\mathcal{A}$ for $N$ intermediate states can be written as
\begin{equation}\label{2eqSumCiFi}
\begin{split}
\mathcal{A}=\sum_j^N c_jF_j(\vec{x}),
\end{split}
\end{equation}
where $c_j$ is a complex coefficient containing the relative contribution of decay channel. $F(\vec{x})$ depicts the dynamics of the intermediate resonance, which is given by~\cite{Aaij:2020ypa, Aaij:2020hon}
\begin{equation}\label{2eqFx}
\begin{split}
F(\vec{x})=R\left(m(D\bar{D})\right)\times T(\vec{p},\vec{q})\times X(|\vec{p}|)\times X(|\vec{q}|).
\end{split}
\end{equation}
In this equation, the $R$ function is a relativistic Breit-Wigner function. The $T$ function describes the angular dependence of $F(\vec{x})$ with Zemach tensor form~\cite{Zemach:1968zz, Zemach:1963bc}, and the $X$ function is a Blatt-Weisskopf barrier form factor. $\vec{p}$ and $\vec{q}$ denote the momentum of the particle produced by the resonance and the momentum of the spectator particle, respectively. The relativistic Breit-Wigner line shape is given as
\begin{equation}\label{2eqRfun}
\begin{split}
R(m)=\frac{1}{(m_0^2-m^2)-im_0\Gamma(m)},
\end{split}
\end{equation}
where $m_0$ is the nominal mass of the resonance and $\Gamma(m)$ denotes the decay width, which is expressed as
\begin{equation}\label{2eqGammafun}
\begin{split}
\Gamma(m)=\Gamma_0\left(\frac{q}{q_0}\right)^{2L+1}\left(\frac{m_0}{m}\right)X^2\left(|q|\right).
\end{split}
\end{equation}
Here, the mass and width are $m_0=3.9238$ GeV and $\Gamma_0=0.0174$ GeV for $\chi_{c0}(2P)$, respectively. The $T$ function and $X$ function employed in Eq.~(\ref{2eqFx}) are given by
\begin{equation}\label{2eqTfun}
\begin{split}
L=0:\quad&T(\vec{p},\vec{q})=1,\\
&X(z)=1,\\
L=1:\quad&T(\vec{p},\vec{q})=-2\vec{p}\cdot\vec{q},\\
&X(z)=\sqrt{\frac{1+z_0^2}{1+z^2}},\\
L=2:\quad&T(\vec{p},\vec{q})=\frac{4}{3}\left[3(\vec{p}\cdot\vec{q})^2-\left(|\vec{p}||\vec{q}|\right)^2\right],\\
&X(z)=\sqrt{\frac{z_0^4+3z_0^2+9}{z^4+3z^2+9}},\\
\end{split}
\end{equation}
where $L$ represents the angular momentum between the resonance and the spectator particle. In the $B^+\to K^+D^+D^-$ process, the angular momentum $L$ is also equal to the spin of the resonance. With the $R$, $T$, and $X$ functions, $F(\vec{x})$ for a resonance in Eq.~(\ref{2eqFx}) can be constructed. Interested readers can find more details of the analysis formalism in Ref.~\cite{Back:2017zqt}.

With the $F(\vec{x})$ function, the fit fraction is defined as
\begin{equation}\label{2eqFitfrac}
\begin{split}
\mathcal{F}_j=\frac{\int |c_jF_j(\vec{x})|^2{\rm d}\vec{x}}{\int |\mathcal{A}|^2{\rm d}\vec{x}},
\end{split}
\end{equation}
which means the integral of the squared amplitude of one resonance divided by the integral of the squared total amplitude. The index $j$ denotes a single resonance. The parameter $c_j$ is fitted from the experimental data, which involves the information of production and decay of the intermediate resonance. In order to extract the relevant branching fraction contained in $c_j$, we can write the amplitude in another form, which contains the amplitudes of the $B\to K\chi_{c0}(2P)$ and $\chi_{c0}(2P)\to D\bar{D}$ processes. With the new form, the amplitude is
\begin{equation}\label{2eqMFM}
\begin{split}
\mathcal{M}^{B\to KD\bar{D}}_{\chi_{c0}(2P)}=\mathcal{M}^{B\to K\chi_{c0}(2P)}\times F(\vec{x})\times \mathcal{M}^{\chi_{c0}(2P)\to D\bar{D}},
\end{split}
\end{equation}
where $F(\vec{x})$ still represents the resonance. $\mathcal{M}^{B\to K\chi_{c0}(2P)}$ and $\mathcal{M}^{\chi_{c0}(2P)\to D\bar{D}}$ are amplitudes for the $B\to K\chi_{c0}(2P)$ and $\chi_{c0}(2P)\to D\bar{D}$ processes, respectively. These two amplitudes can be replaced by their decay width through the decay formula in the narrow width approximation. The decay width of any $A\to BC$ process reads as \cite{Patrignani:2016xqp}
\begin{equation}\label{2eqWid}
\begin{split}
\Gamma(s)=\frac{(2\pi)^4}{2\sqrt{s}}\frac{(2J_B+1)(2J_C+1)}{2J_A+1}g^2\int |\gamma(s)|^2{\rm d}\Phi,
\end{split}
\end{equation}
where $g$ and $\gamma(s)$ are the coupling constant and the corresponding amplitude, respectively. In the narrow width approximation, $\sqrt{s}$ is a fixed physical mass. Hence, the decay width formula can be simplified as
\begin{equation}\label{2eqSimpM}
\begin{split}
g=\frac{1}{\gamma}\sqrt{\frac{M\Gamma}{\rho}}
\end{split}
\end{equation}
with $\rho=\frac{(2J_B+1)(2J_C+1)}{2J_A+1}\frac{(2\pi)^4}{2}\int {\rm d}\Phi$. $\Gamma$ and $M$ are the decay width and the mass of the initial state $A$, respectively. With Eq.~(\ref{2eqSimpM}), the amplitude $\mathcal{M}^{B\to K\chi_{c0}(2P)}$ employed in Eq.~(\ref{2eqMFM}) is
\begin{equation}\label{2eqMBdecay}
\begin{split}
\mathcal{M}^{B\to K\chi_{c0}(2P)}=\sqrt{\frac{M_B\Gamma^{B\to K\chi_{c0}(2P)}}{\rho_{B\to K\chi_{c0}(2P)}}}.
\end{split}
\end{equation}
In Eq.~(\ref{2eqMBdecay}), the two-body phase space is employed with the expression:
\begin{equation}\label{2eqBdecayPhase}
\begin{split}
\rho_{B\to K\chi_{c0}(2P)}&=\frac{(2J_K+1)(2J_{\chi_{c0}}+1)}{2J_B+1}\frac{|\vec{p}_K|}{8\pi M_B}\\
&=\frac{(2J_{\chi_{c0}}+1)|\vec{p}_K|}{8\pi M_B}.\\
\end{split}
\end{equation}
$J_K$, $J_B$, and $J_{\chi_{c0}}$ are total spin for $K$, $B$, and $\chi_{c0}(2P)$ meson, respectively. $M_B$ and $\vec{p}_K$ are the mass of $B$ meson and the momentum of $K$ meson in $B$ rest frame, respectively. The momentum $\vec{p}_K$ can be expressed as $|\vec{p}_K|=\frac{1}{2M_B}\lambda^{1/2}(M_B^2, M_K^2, M_{\chi_{c0}(2P)}^2)$ with $\lambda(a, b, c)=a^2+b^2+c^2-2ab-2bc-2ca$.

By the similar way, $\mathcal{M}^{\chi_{c0}(2P)\to D\bar{D}}$ can also be expressed as
\begin{equation}\label{2eqMchidecay}
\mathcal{M}^{\chi_{c0}(2P)\to D\bar{D}}=\sqrt{\frac{M_{\chi_{c0}(2P)}\Gamma^{\chi_{c0}(2P)\to D\bar{D}}}{\rho_{\chi_{c0}(2P)\to D\bar{D}}}},
\end{equation}
where $\rho_{\chi_{c0}(2P)\to D\bar{D}}$ denotes the phase space. $M_{\chi_{c0}(2P)}$ and $\Gamma^{\chi_{c0}(2P)\to D\bar{D}}$ are the mass and partial decay width of resonance $\chi_{c0}(2P)$, respectively. The phase space in Eq.~(\ref{2eqMchidecay}) is defined as
\begin{equation}\label{2eqMchiphase}
\begin{split}
\rho_{\chi_{c0}(2P)\to D\bar{D}}&=\frac{(2J_D+1)(2J_{\bar{D}}+1)}{2J_{\chi_{c0}}+1}\frac{|\vec{p}^*_D|}{8\pi M_{\chi_{c0}(2P)}}\\
&=\frac{|\vec{p}^*_D|}{(2J_{\chi_{c0}}+1)8\pi M_{\chi_{c0}(2P)}},
\end{split}
\end{equation}
where $J_D$ and $J_{\bar{D}}$ are the total angular momentum of $D$ and $\bar{D}$ states, respectively.

With the above preparation, the total amplitude of the $B\to KD\bar{D}$ process in Eq.~(\ref{2eqMFM}) has the expression
\begin{equation}\label{2eqAMPBKDD}
\begin{split}
\mathcal{M}^{B\to KD\bar{D}}_{\chi_{c0}(2P)}(\vec{x})=F(\vec{x})\times\sqrt{\frac{M_BM_{\chi_{c0}(2P)}\Gamma^{B\to K\chi_{c0}(2P)}\Gamma^{\chi_{c0}\to D\bar{D}}}{\rho_{B\to K\chi_{c0}(2P)}\rho_{\chi_{c0}(2P)\to D\bar{D}}}},
\end{split}
\end{equation}
where the partial decay width $\Gamma^{B\to K\chi_{c0}(2P)}$ equals $\mathcal{B}[B\to K\chi_{c0}(2P)]\times \Gamma_B$. Here, the construction of the amplitude in Eq.~(\ref{2eqAMPBKDD}) is inspired by the method employed to calculate the process of electron-positron collisions in Refs.~\cite{Ablikim:2020cyd, Ablikim:2019apl, Ablikim:2016qzw, Jia:2020epr, BESIII:2020pov}.

With the amplitude in Eq.~(\ref{2eqAMPBKDD}), the fit fraction can also be written as
\begin{equation}\label{2eqFitfrac2}
\begin{split}
\mathcal{F}_j=\frac{\int \left|\mathcal{M}_j(\vec{x})\right|^2{\rm d}\vec{x}}{\int \left|\mathcal{A}\right|^2{\rm d}\vec{x}},
\end{split}
\end{equation}
where the three-body phase space is integrated in the resonance rest frame. With the above fit fraction, the ratio of different fit fractions can be derived as
\begin{equation}\label{2eqRatio}
\begin{split}
\frac{\mathcal{F}_{\chi_{c0}(2P)}}{\mathcal{F}_{\psi(3770)}}&=\frac{\int\left|c_{\chi_{c0}(2P)}F_{\chi_{c0}(2P)}(\vec{x})\right|^2{\rm d}\vec{x}}
{\int\left|c_{\psi(3770)}F_{\psi(3770)}(\vec{x})\right|^2{\rm d}\vec{x}}\\
&=\frac{\int \left|\mathcal{M}_{\chi_{c0}(2P)}^{B\to KD\bar{D}}(\vec{x})\right|^2{\rm d}\vec{x}}{\int \left|\mathcal{M}_{\psi(3770)}^{B\to KD\bar{D}}(\vec{x})\right|^2{\rm d}\vec{x}}.\\
\end{split}
\end{equation}
The ratio $\frac{\mathcal{F}_{\chi_{c0}(2P)}}{\mathcal{F}_{\psi(3770)}}$ in the first line has already been measured in Refs.~\cite{Aaij:2020ypa, Aaij:2020hon}. And the integral $\int|F_j(\vec{x})|^2{\rm d}\vec{x}$ is normalized. Since $\mathcal{B}[B\to K\psi(3770)]$ had been determined in many experimental measurements \cite{Aubert:2005vi, Abe:2003zv}, the branching fraction $\mathcal{B}[B\to K\chi_{c0}(2P)]$ can be derived with the ratio in Eq.~(\ref{2eqRatio}). With the explicit form of the amplitude in Eq.~(\ref{2eqAMPBKDD}), the branching fraction is
\begin{equation}\label{2eqBr}
\small
\begin{split}
&\mathcal{B}\left[B\to K\chi_{c0}(2P)\right]=\frac{\mathcal{F}_{\chi_{c0}(2P)}}{\mathcal{F}_{\psi(3770)}}\\
&\times\frac{M_{\psi(3770)}\mathcal{B}[B\to K\psi(3770)]\mathcal{B}[\psi(3770)\to D\bar{D}]}{M_{\chi_{c0}(2P)}\mathcal{B}[\chi_{c0}(2P)\to D\bar{D}]}\\
&\times\frac{\Gamma_{\psi(3770)}\rho_{B\to K\chi_{c0}(2P)}\rho_{\chi_{c0}(2P)\to D\bar{D}}}{\Gamma_{\chi_{c0}(2P)}\rho_{B\to K\psi(3770)}\rho_{\psi(3770)\to D\bar{D}}}.\\
\end{split}
\end{equation}

To determine the branching ratio $\mathcal{B}[B\to K\chi_{c0}(2P)]$, many different physical quantities in Eq.~(\ref{2eqBr}) should be adopted as input. We collect these values with the error bars in Table~\ref{errorbars}, which are helpful to estimate the uncertainty of $\mathcal{B}[B\to K\chi_{c0}(2P)]$.

\begin{table}[htbp]
\scriptsize
\caption{The values as input in Eq.~(\ref{2eqBr}).}
\label{errorbars}
\renewcommand\arraystretch{1.6}
\begin{threeparttable}
\begin{tabular*}{86mm}{@{\extracolsep{\fill}}ccc}
\toprule[1.0pt]
\toprule[1.0pt]
I&$\mathcal{B}[\chi_{c0}(2P)\to D\bar{D}]$=90\%&\\
\toprule[0.8pt]
II&$\mathcal{B}[\psi(3770)\to D\bar{D}]=0.93\pm0.09$&\\
&$\mathcal{B}[B\to K\psi(3770)]=(4.9\pm1.3)\times10^{-4}$&\\
\toprule[0.8pt]
{III}&$M_{\chi_{c0}(2P)}=3.9238\pm0.0015\pm0.0004$ GeV & $\mathcal{F}_{\psi(3770)}=14.5\pm1.2\pm0.8$\\
&$\Gamma_{\chi_{c0}(2P)}=0.0174\pm0.0051\pm0.0008$ GeV & $\mathcal{F}_{\chi_{c0}(2P)}=3.7\pm0.9\pm0.2$\\
&$M_{\psi(3770)}=3.7781\pm0.0009$ GeV&$\Gamma_{\psi(3770)}=0.0272\pm0.0010$ GeV\\
\bottomrule[1pt]
\bottomrule[1pt]
\end{tabular*}
\begin{tablenotes}
\item[I] {This value is a theoretical estimate.
Although the OZI-favored mode $\chi_{c0}(2P)\to D\bar{D}$ is dominant,
these OZI-suppressed modes of $\chi_{c0}(2P)$ have considerable contribution to the total width of  $\chi_{c0}(2P)$. As
shown in this work, a branching ratio $\chi_{c0}(2P)\to J/\psi \omega$ of $~ 3\%$ can be obtained (see Eq. (\ref{4eqBrchiomepsi})). Additionally, there exist other similar OZI-suppressed modes like $\chi_{c0}(1P)\pi\pi$, $\chi_{c2}(1P)\pi\pi$, $\chi_{c1}(1P)\eta$, and $\eta_c\eta$.
If making a qualitatively estimate of these
modes from the experimental measurements in $\psi(3770)$ \cite{BES:2003bes, CLEO:2005zky}, then the total branching ratio of all OZI-suppressed modes might reach $(5-10)\%$, and consequently an
estimate for $\mathcal{B}[\chi_{c0}(2P) \to D\bar D]$ smaller than one.
In this work, we take $\mathcal{B}[\chi_{c0}(2P) \to D\bar D]=90\%$ as input.}
\item[II] The branching fractions are listed in the PDG~\cite{Zyla:2020zbs}.
\item[III] These values are employed in the experimental analysis in Ref.~\cite{Aaij:2020ypa}.
\end{tablenotes}
\end{threeparttable}
\end{table}

In Eq.~(\ref{2eqBr}), it is obvious that the error bar of $\mathcal{B}[B\to K\chi_{c0}(2P)]$ is propagated from the uncertainty of physical quantities listed in Table~\ref{errorbars}, where the error transfer formula will be employed and has the form,
\begin{equation}\label{2eqerrortran}
\begin{split}
u_f=\sqrt{\left(\frac{\partial F}{\partial x}u_x\right)^2+\left(\frac{\partial F}{\partial y}u_y\right)^2+\left(\frac{\partial F}{\partial z}u_z\right)^2},
\end{split}
\end{equation}
where $x$, $y$, and $z$ are the independent variables, and $u_x$, $u_y$, and $u_z$ are their uncertainties. The variable $f$ is determined by the function $f=F(x, y, z)$, so the uncertainty $u_f$ can be calculated by this formula.

Based on Eq.~(\ref{2eqerrortran}), the uncertainty of $\mathcal{B}[B\to K\chi_{c0}(2P)]$ is also obtained. The final result of $\mathcal{B}[B\to K\chi_{c0}(2P)]$ is
\begin{equation}\label{2eqBrB2Kchi}
\begin{split}
\mathcal{B}[B\to K\chi_{c0}(2P)]=(3.5\pm1.4)\times10^{-4}.
\end{split}
\end{equation}
In the above estimate, this branching fraction is actually determined by the scaling point $\mathcal{B}[B\to K\psi(3770)]$.
In the next section, the extracted value of $\mathcal{B}[B\to K\chi_{c0}(2P)]$ will be reproduced by introducing the rescattering mechanism.

\section{Understanding the extracted branching ratio of $B\to K\chi_{c0}(2P)$ via the rescattering mechanism}\label{sec3}

The rescattering mechanism has been widely used in the decay studies of charmonium and charmoniumlike states for a long time, which indicates the obvious effect on the resulting branching ratio \cite{Liu:2009dr, Liu:2006df, Liu:2009iw, Liu:2006dq}. The charmonium production in $B$ meson decay is another kind of process influenced by the rescattering mechanism. For example, the branching fraction of  $\mathcal{B}(B\to K X(3823))$ was explained well by treating $X(3823)$ as charmonium $\psi_2(1^3D_2)$ in this mechanism and $\mathcal{B}(B\to K\eta_{c2}(1^1D_2))$ and $\mathcal{B}(B\to K\psi_3(1^3D_3))$ are predicted in Ref.~\cite{Xu:2016kbn}.
In Refs.~\cite{Colangelo:2003sa, Colangelo:2002mj}, the authors have been studied the $B\to K\chi_{c0}(1P)$ and $B\to Kh_c(1P)$ processes. For $B\to K\chi_{c0}(1P)$, the rescattering mechanism \cite{Colangelo:2002mj} was adopted to avoid the difficulty when applying the na\"ive factorization approach \cite{Buchalla:1995vs}
to depict such decay \cite{Zyla:2020zbs}.

\begin{figure}[htbp]
\centering
\includegraphics[width=8.6cm,keepaspectratio]{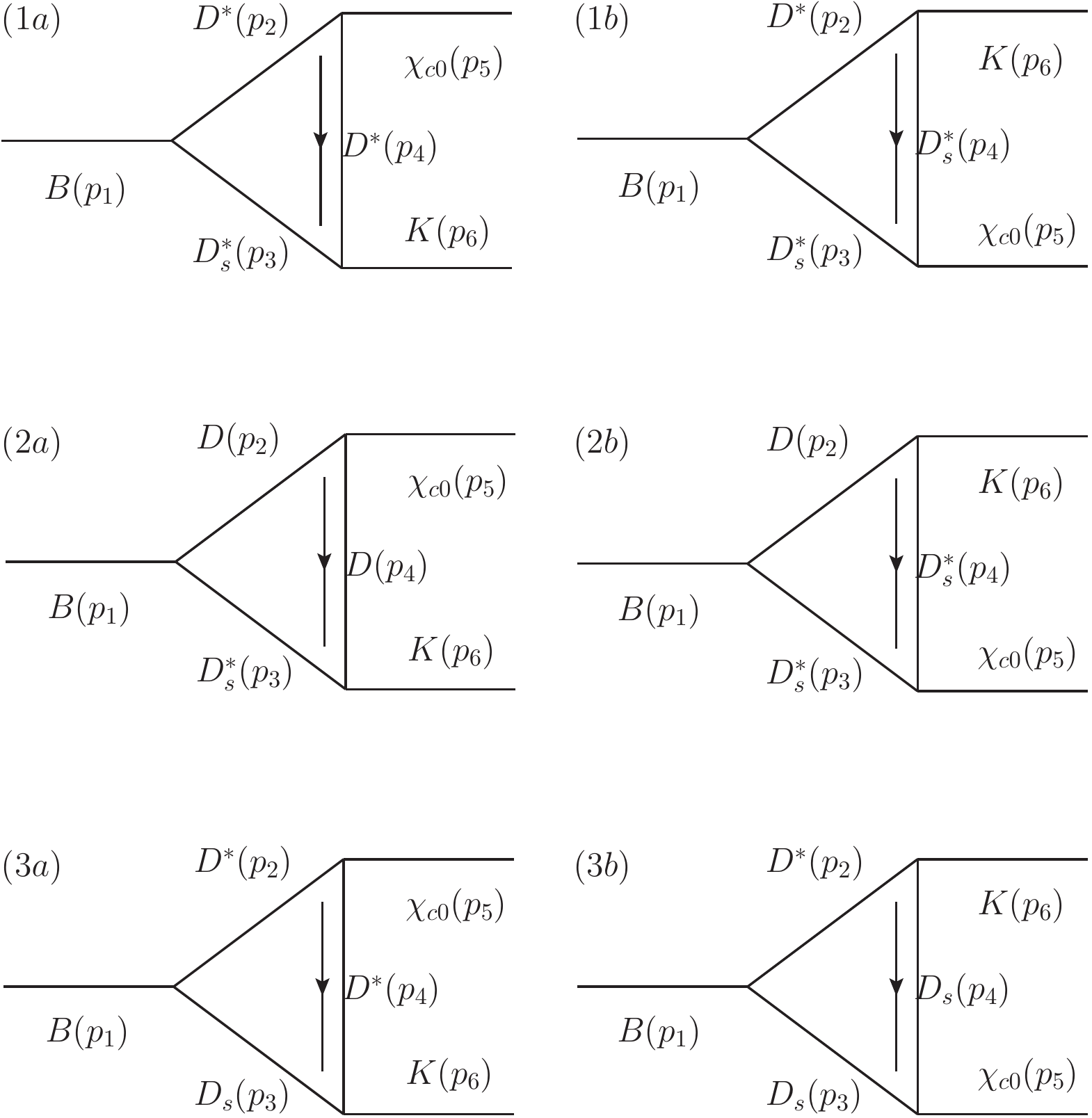}
\caption{Feynman diagrams for the $B\to K\chi_{c0}(2P)$ processes via the rescattering mechanism.}
\label{FigFeyn}
\end{figure}

In the rescattering mechanism for the discussed $B\to K\chi_{c0}(2P)$ process, a $B$ meson decays to a pair of charmed and charmed-strange meson firstly, and then via the rescattering, the $D^{(*)}$ and $D_s^{(*)}$ meson are transferred into final state $\chi_{c0}(2P)$ and $K$. The Feynman diagrams depicting the $B\to K\chi_{c0}(2P)$ process are shown in Fig.~\ref{FigFeyn}.

The weak interaction vertex of the $B\to D^{(*)}D_s^{(*)}$ process is included in Fig.~\ref{FigFeyn}, and the effective weak Hamiltonian is written as
\begin{equation}\label{3eqWH}
\begin{split}
H_W&=\frac{G_F}{\sqrt{2}}[V_{cb}V_{cs}^*(c_1(\mu)\mathcal{O}_1(\mu)+c_2(\mu)\mathcal{O}_2(\mu))\\
&\quad-V_{tb}V_{ts}^*\sum_{i=3}^{10}c_i(\mu)\mathcal{O}_i(\mu)]+h.c.,\\
\end{split}
\end{equation}
where $c_i$ is the Wilson coefficient, and $\mathcal{O}_i$ denotes the operator.
Ignoring the small contributions from $\mathcal{O}_3\sim\mathcal{O}_{10}$, the amplitude of $B\to D^{(*)}D_s^{(*)}$ is
\begin{equation}\label{3eqWHME}
\begin{split}
\langle D^{(*)}D_s^{(*)}|H_W|B\rangle=&\frac{G_F}{\sqrt{2}}V_{cb}V_{cs}^*a_1\langle D^{(*)}|V^{\mu}-A^{\mu}|B\rangle\\
&\times\langle D_s^{(*)}|V_{\mu}-A_{\mu}|0\rangle\\
\end{split}
\end{equation}
with $a_1=c_1+\frac{c_2}{N_c}$. With the Isgur-Wise function, the matrix elements employed in Eq.~(\ref{3eqWHME}) have the form
\begin{equation}
\begin{split}
\langle D(v^\prime)|V^{\mu}|B(v)\rangle&=\sqrt{m_Bm_D}\xi(v\cdot v^\prime)(v^\prime+v)^\mu,\\
\langle D^*(v^\prime,\epsilon)|V_\mu|B(v)\rangle&=i\sqrt{m_Bm_{D^*}}\xi(v\cdot v^\prime)\varepsilon_{\mu\nu\alpha\beta}\epsilon^{\nu*}v^{\prime\alpha}v^{\beta},\\
\langle D^*(v^\prime,\epsilon)|A^{\mu}|B(v)\rangle=&\sqrt{m_Bm_{D^*}}\xi(v\cdot v^\prime)\\
&\left((1+v\cdot v^\prime)g^{\alpha\mu}-v^\alpha v^{\prime\mu}\right)\epsilon^{*}_\alpha,\\
\langle0|A^\mu|D_s(v)\rangle&=f_{D_s}m_{D_s}v^{\mu},\\
\langle0|V^\mu|D^*_s(v,\epsilon)\rangle&=f_{D_s^*}m_{D_s^*}\epsilon^{\mu}.\\
\end{split}
\end{equation}
where $v^{\mu}$ is the velocity of the corresponding heavy meson and $\xi(v\cdot v^{\prime})$ represents the Isgur-Wise function. With the above matrix elements, the amplitude of the Feynman diagrams for the $B\to D^{(*)}D_s^{(*)}$ part can be given by
\begin{equation}
\begin{split}
\langle D(p_2)D_s(p_3)|B(v_1)\rangle&=\frac{G_F}{\sqrt{2}}V_{cb}V_{cs}^*a_1\sqrt{m_Bm_D}\xi(v\cdot v^\prime)\\
&\times\left(\frac{p_2^\mu}{m_D}+v^\mu_1\right)f_{D_s}p_3,\\
\langle D^*(p_2)D_s^*(p_3)|B(v_1)\rangle&=\frac{G_F}{\sqrt{2}}V_{cb}V_{cs}^*a_1\sqrt{m_Bm_{D^*}}\xi(v\cdot v^\prime)\\
&\times\left(i\varepsilon_{\mu\nu\alpha\beta}\frac{p_2^\alpha}{m_{D^*}}v_1^\beta-(1+\omega)g_{\mu\nu}+v_{1\nu}\frac{p_{2\mu}}{m_{D^*}}\right)\\
&\times\epsilon_2^{*\nu}f_{D_s^*}m_{D_s^*}\epsilon_{D_s^*}^{*\mu},\\
\langle D(p_2)D_s^*(p_3)|B(v_1)\rangle&=\frac{G_F}{\sqrt{2}}V_{cb}V_{cs}^*a_1\sqrt{m_Bm_{D}}\xi(v\cdot v^\prime)\\
&\times\left(\frac{p_2^\mu}{m_D}+v_1^\mu\right)f_{D_s^*}m_{D_s^*}\epsilon_{3\mu}^*,\\
\langle D^*(p_2)D_s(p_3)|B(v_1)\rangle&=\frac{G_F}{\sqrt{2}}V_{cb}V_{cs}^*a_1\sqrt{m_Bm_{D^*}}\xi(v\cdot v^\prime)\\
&\times\left(i\varepsilon_{\mu\nu\alpha\beta}\frac{p_2^\alpha}{m_{D^*}}v_1^\beta-(1+\omega)g_{\mu\nu}+v_{1\nu}\frac{p_{2\mu}}{m_{D^*}}\right)\\
&\times\epsilon_2^{*\nu}f_{D_s}p_3^\mu.\\
\end{split}
\end{equation}

For the $D^{(*)}D_s^{(*)}K$ vertex, the interaction is considered in heavy quark effective theory, then the compact Lagrangian for the vertex is \cite{Wise:1992hn}
\begin{equation}\label{3eqcomLag}
\begin{split}
\mathcal{L}=ig{\rm Tr}[H_b\gamma_\mu\gamma_5\mathcal{A}^\mu_{ba}\bar{H}_a],
\end{split}
\end{equation}
where $\mathcal{A}^\mu_{ba}$ is an axial vector current, which can be expanded as $\mathcal{A}^\mu_{ba}=\frac{i}{f_\pi}\partial^\mu \mathcal{M}_{ba}+\cdots$. Here, $\mathcal{M}_{ba}$ represents the octet of pseudoscalar mesons, where the field of $K$ meson is contained. $H$ represents the super field of heavy-light meson which can be constructed as $H_a=(\frac{1+\slashed{v}}{2})(D_a^{*\mu}\gamma_\mu+iD_a\gamma_5)$ and $\bar{H}_a=\gamma_0H_a^\dag\gamma_0$.

Expanding the compact Lagrangian in Eq.~(\ref{3eqcomLag}), the effective Lagrangians for the interaction of $D^{(*)}D_s^{(*)}K$ are given as
\begin{equation}
\begin{split}
\mathcal{L}_{D_s^*DK}&=ig_{D_s^*DK}\bar{D}_{s\mu}^{*}D\partial^{\mu}K,\\
\mathcal{L}_{D^*D_sK}&=-ig_{D_sD^*K}\bar{D}_sD_\mu^*\partial^\mu K,\\
\mathcal{L}_{D^*D_s^*K}&=g_{D_s^*D^*K}\varepsilon_{\mu\nu\alpha\beta}\partial^\mu K\bar{D}_s^{*\nu}\partial^{\alpha}D^{*\beta},\\
\end{split}
\end{equation}
where the coupling constants are defined as $g_{DD_s^*K}=g_{D^*D_sK}=\sqrt{m_{D_s^*}m_D}(\frac{2g}{f_\pi})$ and $g_{D^*D_s^*K}=\frac{\sqrt{m_{D_s^*}m_{m_{D^*}}}}{m_{D^*}}(\frac{2g}{f_\pi})$. The decay constant $f_\pi$ is equal to 132 MeV.

The coupling of a $P-$wave charmonium and charmed meson pair is indicated as the compact Lagrangian, i.e.,
\begin{equation}
\mathcal{L}=ig_1{\rm Tr}[\mathcal{P}^{\mu}\bar{H}_{\bar{Q}}\gamma_{\mu}\bar{H}_Q].\\
\end{equation}
In this Lagrangian, $\mathcal{P}^{\mu}$ represents the multiplet of $P-$wave charmonia, which has the following expression:
\begin{eqnarray}
\mathcal{P}^\mu&=&\frac{1+\slashed{v}}{2}\Bigg(\chi_{c2}^{\mu\alpha}\gamma_\alpha+\frac{1}{\sqrt{2}}\varepsilon^{\mu\delta\alpha\beta}v_\delta\gamma_{\alpha}\chi_{c1\beta}+\frac{1}{\sqrt{3}}
(\gamma^{\mu}-v^\mu)\chi_{c0}\nonumber\\&&
+h_1^\mu\gamma_5\Bigg)\frac{1-\slashed{v}}{2}.
\end{eqnarray}

The effective Lagrangians that represent the interaction of a $\chi_{c0}(2P)$ state and charmed meson pair are expanded from the above compact Lagrangian, which can be written as
\begin{equation}
\begin{split}
\mathcal{L}_{\chi_{c0}(2P)D_{(s)}\bar{D}_{(s)}}&=ig_{\chi_{c0}(2P)D_{(s)}\bar{D}_{(s)}}\chi_{c0}D_{(s)}\bar{D}_{(s)},\\
\mathcal{L}_{\chi_{c0}(2P)D^*_{(s)}\bar{D}^*_{(s)}}&=ig_{\chi_{c0}(2P)D^*_{(s)}\bar{D}^*_{(s)}}\chi_{c0}D^*_{(s)\mu}\bar{D}^{*\mu}_{(s)}.\\
\end{split}
\end{equation}
The Lagrangians $\mathcal{L}_{\chi_{c0}(2P)D_sD_s}$ and $\mathcal{L}_{\chi_{c0}(2P)D_s^*D_s^*}$ have a similar expression with $\mathcal{L}_{\chi_{c0}(2P)D\bar{D}}$ and $\mathcal{L}_{\chi_{c0}(2P)D^*\bar{D}^*}$, respectively.
Here the coupling constants of the $\chi_{c0}(2P)$ state related by the original coupling constant $g_1$ are
\begin{equation}\label{3eqCoupCon}
\begin{split}
g_{\chi_{c0}(2P)D_{(s)}\bar{D}_{(s)}}&=2\sqrt{3}g_1\sqrt{m_{\chi_{c0}(2P)}m_{D_{(s)}}m_{\bar{D}_{(s)}}},\\
g_{\chi_{c0}(2P)D^*_{(s)}\bar{D}^*_{(s)}}&=\frac{2}{\sqrt{3}}g_1\sqrt{m_{\chi_{c0}(2P)}m_{D^*_{(s)}}m_{\bar{D}^*_{(s)}}},\\
\end{split}
\end{equation}
where $g_{\chi_{c0}(2P)D_s\bar{D}_s}$ and $g_{\chi_{c0}(2P)D_s^*\bar{D}_s^*}$ are the coupling constants for $\mathcal{L}_{\chi_{c0}(2P)D_sD_s}$ and $\mathcal{L}_{\chi_{c0}(2P)D_s^*D_s^*}$, respectively.

With the above effective Lagrangians, the amplitudes $\mathcal{M}(D^{(*)}D_s^{(*)}\to K\chi_{c0}(2P))$ are obtained. Combined with the amplitudes of $B$ decay processes, we write out the imaginary part of amplitudes for the $B\to K\chi_{c0}(2P)$ process by the Cutkosky cutting rule
\begin{equation}
\begin{split}
&Abs(\mathcal{M}_{B\to K\chi_{c0}(2P)})\\
&=\frac{|p_2|}{32\pi^2m_B}\int {\rm d}\Omega_{p_2}\mathcal{M}_{B\to D^{(*)}D_s^{(*)}}\mathcal{M}_{D^{(*)}D^{(*)}_s\to K\chi_{c0}(2P)}.\\
\end{split}
\end{equation}
The concrete expressions of decay amplitudes corresponding to Fig.~\ref{FigFeyn} are written as
\begin{equation}\label{3eqamp1}
\begin{split}
&Abs(\mathcal{M}_{1a})\\
=&\frac{|p_2|}{32\pi^2m_B}\int {\rm d}\Omega \frac{G_F}{\sqrt{2}}V_{cb}V_{cs}^*a_1\sqrt{m_Bm_{D^*}}\,\xi(\omega)\\
&\times\left(i\varepsilon_{\gamma\delta\alpha\beta}\frac{p_2^\alpha}{m_{D^*}}v_1^\beta-(1+\omega)g_{\gamma\delta}+v_{1\delta}\frac{p_{2\gamma}}{m_{D^*}}\right)f_{D_s^*}m_{D_s^*}\\
&\times(-i)g_{D^*D_s^*K}g_{\chi_{c0}D^*\bar{D}^*}\varepsilon_{\mu\nu\tau\rho}p_6^\mu p_4^\tau \left (-g^\delta_\theta+\frac{p_{2\theta}p_2^\delta}{m_{D^*}^2}\right)\\
&\times\left(-g^{\nu\gamma}+\frac{p_3^\nu p_3^\gamma}{m_{D_s^*}^2}\right)\left(-g^{\rho\theta}+\frac{p_4^\rho p_4^\theta}{m_{D^*}^2}\right)\frac{1}{p_4^2-m_{D^*}^2}\mathfrak{F}^2(p_4^2,\Lambda^2),\\
\end{split}
\end{equation}
\begin{equation}\label{3eqamp2}
\begin{split}
&Abs(\mathcal{M}_{1b})\\
=&\frac{|p_2|}{32\pi^2m_B}\int {\rm d}\Omega \frac{G_F}{\sqrt{2}}V_{cb}V_{cs}^*a_1\sqrt{m_Bm_{D^*}}\,\xi(\omega)\\
&\times\left(i\varepsilon_{\gamma\delta\alpha\beta}\frac{p_2^\alpha}{m_{D^*}}v_1^\beta-(1+\omega)g_{\gamma\delta}+v_{1\delta}\frac{p_{2\gamma}}{m_{D^*}}\right)f_{D_s^*}m_{D_s^*}\\
&\times(-i)g_{D^*D_s^*K}g_{\chi_{c0}D^*\bar{D}^*}\varepsilon_{\mu\nu\tau\rho}p_6^\mu p_2^\tau \left(-g^{\rho\delta}+\frac{p_2^\rho p_2^\delta}{m_{D^*}^2}\right)\\
&\times\left(-g^{\theta\gamma}+\frac{p_3^\theta p_3^\gamma}{m_{D_s^*}^2}\right)\left(-g_\theta^\nu+\frac{p_{4\theta} p_4^\nu}{m_{D_s^*}^2}\right)\frac{1}{p_4^2-m_{D_s^*}^2}\mathfrak{F}^2(p_4^2,\Lambda^2),\\
\end{split}
\end{equation}
\begin{equation}\label{3eqamp3}
\begin{split}
&Abs(\mathcal{M}_{2a})\\
=&\frac{|p_2|}{32\pi^2m_B}\int {\rm d}\Omega \frac{G_F}{\sqrt{2}}V_{cb}V_{cs}^*a_1\sqrt{m_Bm_D}\,\xi(\omega)\\
&\times(-i)g_{DD_s^*K}g_{\chi_{c0}D\bar{D}}p_6^\mu f_{D_s^*}m_{D_s^*}\\
&\times\left(v_{1\gamma}+\frac{p_{2\gamma}}{m_D}\right)\left(-g_\mu^\gamma+\frac{p_{3\mu}p_3^\gamma}{m_3^2}\right)\frac{1}{p_4^2-m_D^2}\mathfrak{F}^2(p_4^2,\Lambda^2),\\
\end{split}
\end{equation}

\begin{equation}\label{3eqamp4}
\begin{split}
&Abs(\mathcal{M}_{2b})\\
=&\frac{|p_2|}{32\pi^2m_B}\int {\rm d}\Omega \frac{G_F}{\sqrt{2}}V_{cb}V_{cs}^*a_1\sqrt{m_Bm_D}\,\xi(\omega)\\
&\times\left(v_{1\gamma}+\frac{p_{2\gamma}}{m_D}\right)f_{D_s^*}m_{D_s^*}i g_{DD_s^*K}g_{\chi_{c0}D_s^*\bar{D}_s^*}p_6^\mu\\
&\times\left(-g^{\alpha\gamma}+\frac{p_3^\alpha
p_3^\gamma}{m_3^2}\right)\left(-g_{\alpha\mu}+\frac{p_{4\alpha}p_{4\mu}}{m_4^2}\right)\frac{1}{p_4^2-m_{D_s^*}^2}\mathfrak{F}^2(p_4^2,\Lambda^2),\\
\end{split}
\end{equation}
\begin{equation}\label{3eqamp5}
\begin{split}
&Abs(\mathcal{M}_{3a})\\
=&\frac{|p_2|}{32\pi^2m_B}\int {\rm d}\Omega \frac{G_F}{\sqrt{2}}V_{cb}V_{cs}^*a_1\sqrt{m_Bm_D^*}\,\xi(\omega)\\
&\times\left(i\varepsilon_{\gamma\delta\alpha\beta}\frac{p_2^\alpha}{m_{D^*}}v_1^\beta-(1+\omega)g_{\gamma\delta}+v_{1\delta}\frac{p_{2\gamma}}{m_{D^*}}\right)f_{D_s}\\
&\times(-i)g_{D^*D_sK}g_{\chi_{c0}D^*\bar{D}^*}p_6^\mu p_3^\gamma\\
&\times\left(-g^\delta_\nu+\frac{p_{2\nu}p_2^\delta}{m_{D^*}^2}\right)\left(-g_\mu^\nu+\frac{p_{4\mu}p_4^\nu}{m_{D^*}^2}\right)\frac{1}{p_4^2-m_{D^*}^2}\mathfrak{F}^2(p_4^2,\Lambda^2),\\
\end{split}
\end{equation}
\begin{equation}\label{3eqamp6}
\begin{split}
&Abs(\mathcal{M}_{3b})\\
=&\frac{|p_2|}{32\pi^2m_B}\int {\rm d}\Omega \frac{G_F}{\sqrt{2}}V_{cb}V_{cs}^*a_1\sqrt{m_Bm_D^*}\,\xi(\omega)\\
&\times\left(i\varepsilon_{\gamma\delta\alpha\beta}\frac{p_2^\alpha}{m_{D^*}}v_1^\beta-(1+\omega)g_{\gamma\delta}+v_{1\delta}\frac{p_{2\gamma}}{m_{D^*}}\right)f_{D_s}\\
&\times ig_{D^*D_sK}g_{\chi_{c0}D_sD_s}p_6^\mu p_3^\gamma\left(-g^\delta_\mu+\frac{p_{2\mu}p_2^\delta}{m_{D^*}^2}\right)\frac{1}{p_4^2-m_{D_s}^2}\mathfrak{F}^2(p_4^2,\Lambda^2).\\
\end{split}
\end{equation}
Here, $\mathfrak{F}(p^2,\Lambda^2)$ is the monopole form factor \cite{Cheng:2004ru}
\begin{equation}\label{3eqFormFactor}
\begin{split}
\mathfrak{F}(p^2,\Lambda^2)=\frac{\Lambda^2-m^2}{\Lambda^2-p^2},
\end{split}
\end{equation}
where $m$ represents the mass corresponding to an exchanged charmed meson with momentum $p$, and $\Lambda$ is a cutoff parameter. Generally, the cutoff parameter can be parametrized as $\Lambda=m+\alpha\Lambda_{QCD}$ where $\Lambda_{QCD}=220$ MeV~\cite{Cheng:2004ru}.

The total amplitude is
\begin{equation}\label{3eqtotamp}
\begin{split}
\mathcal{M}^{tot}_{B\to K\chi_{c0}(2P)}=\sum_iAbs(\mathcal{M}_i),
\end{split}
\end{equation}
where $i$ denotes the amplitudes in Eqs.~(\ref{3eqamp1})-(\ref{3eqamp6}). Finally, the branching ratio of $B\to K\chi_{c0}(2P)$ process reads as
\begin{equation}\label{3eqBr}
\begin{split}
\mathcal{B}[B\to K\chi_{c0}(2P)]=\frac{1}{\Gamma_B}\frac{1}{8\pi}\frac{|p_K|}{m_B^2}|\mathcal{M}^{tot}_{B\to K\chi_{c0}(2P)}|^2.
\end{split}
\end{equation}
Here, $|p_K|$ and $\Gamma_B$ are the momentum of $K$ meson and total decay width of $B$ meson, respectively.

The input parameters include $G_F=1.16638\times10^{-5}$, $a_1=1$, $V_{cb}=0.04$, and $V_{cs}=1$ \cite{Zyla:2020zbs}, and the strong coupling constant $g$ is equal to $g=0.59$ \cite{Chen:2013yxa}. The decay constant of the $D_s$ meson is measured to be $f_{D_s}=252.9\pm3.7(\rm stat)\pm3.6(\rm syst)$ MeV in 2019 by BESIII \cite{Ablikim:2018jun}. Since there is no measurement for $f_{D_s^*}$, we assume $f_{D^*_s}$=$f_{D_s}=0.2529$ GeV. The expression of the Isgur-Wise function is \cite{Cheng:2003sm}
\begin{equation}\label{3eqISWise}
\begin{split}
\xi(\omega)=1-1.22(\omega-1)+0.85(\omega-1)^2,
\end{split}
\end{equation}
with $\omega=v\cdot v^\prime$.
The resonance parameters of $\chi_{c0}(2P)$ and $\chi_{c2}(2P)$ are measured precisely by LHCb, i.e., $\Gamma_{\chi_{c0}(2P)}=17.4\pm5.1\pm0.8$ MeV and $\Gamma_{\chi_{c2}(2P)}=34.2\pm6.6\pm1.1$ MeV. Since the $D\bar{D}$ channel is the only OZI-allowed channel for $\chi_{c0}(2P)$ state, $\Gamma_{\chi_{c0}(2P)}$ can be used to determine the coupling constant $g_{\chi_{c0}(2P)D\bar{D}}=2.323$ directly. But, for the other coupling constants like $g_{\chi_{c0}(2P)D^*\bar{D}^*}$ and $g_{\chi_{c0}(2P)D_s^{(*)}D_s^{(*)}}$, there is no direct experimental information to limit their values. To determine these coupling constants, we firstly determine $g_1=21.87$ in compact Lagrangian with $\Gamma_{\chi_{c2}(2P)}=34.2$ MeV. Then the coupling constants $g_{\chi_{c0}(2P)D^*\bar{D}^*}$ and $g_{\chi_{c0}(2P)D_s^{(*)}\bar{D}_s^{(*)}}$ can be determined according to the relation shown in Eq.~(\ref{3eqCoupCon}).

\begin{figure}[htbp]
\centering
\includegraphics[width=8.6cm,keepaspectratio]{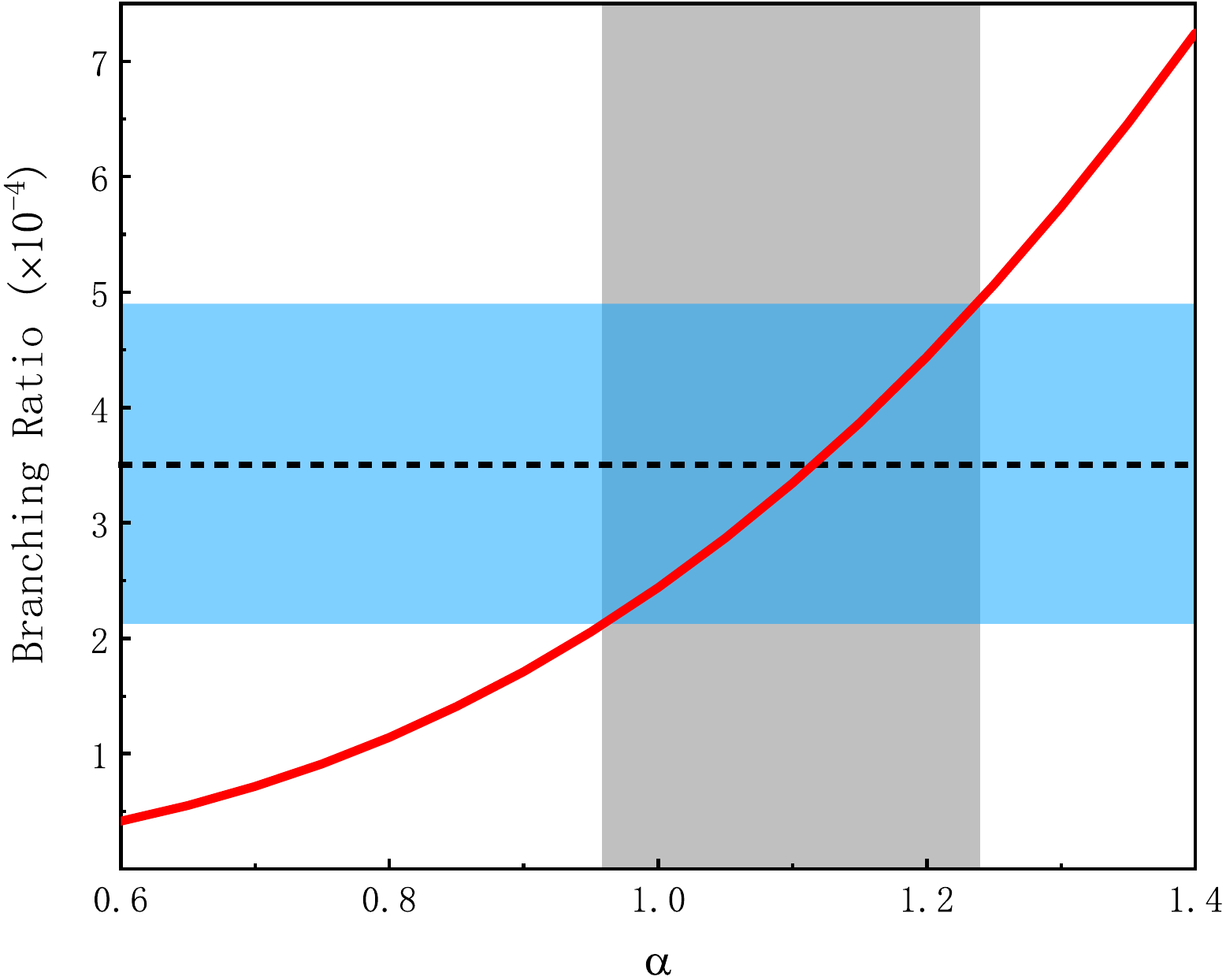}
\caption{The comparison of our result and experimental data for
$\mathcal{B}[B\to K\chi_{c0}(2P)]$. }
\label{FigBr}
\end{figure}

Now, the only unknown parameter in the calculation is the $\alpha$ in Eq.~(\ref{3eqFormFactor}), which makes the branching ratio $\mathcal{B}[B\to K\chi_{c0}(2P)]$ depending only on the choice of $\alpha$. Generally, the cutoff parameter $\Lambda$ is not far from the mass of the exchanged charmed or charmed-strange meson, and $\alpha$ is taken to be of the order of unity \cite{Cheng:2004ru}. In Fig.~\ref{FigBr}, we present the result of the branching ratio dependent on $\alpha$, where the blue block represents a value range of $\mathcal{B}[B\to K\chi_{c0}(2P)]=(3.5\pm1.4)\times10^{-4}$, which is extracted by experimental data as given in the above section, and the red curve is the calculated branching ratio. The gray block shows a range of $\alpha$, where the results from the rescattering mechanism can overlap with the extracted experimental data of $\mathcal{B}[B\to K\chi_{c0}(2P)]$.
Finally, the range of parameter $\alpha$ is found to be $0.95<\alpha<1.24$, which shows that
$\mathcal{B}[B\to K\chi_{c0}(2P)]=(3.5\pm1.4)\times10^{-4}$ can be reproduced well under the rescattering mechanism.
To some extent, assigning the $\chi_{c0}(3930)$ observed by LHCb to be a $\chi_{c0}(2P)$ charmonium is tested.

\section{Discussion and conclusion}\label{sec4}

As shown in Sec. \ref{sec2},
we extract the branching ratio $\mathcal{B}[B\to K\chi_{c0}(2P)]$ (see Eq.~(\ref{2eqBrB2Kchi})), by which we may estimate the product branching fraction $\mathcal{B}[B\to K\chi_{c0}(2P)]\times\mathcal{B}[\chi_{c0}(2P)\to \omega J/\psi]$, where we need the information of the branching ratio $\mathcal{B}[\chi_{c0}(2P)\to \omega J/\psi]$.
We notice the measured result of $\Gamma_{X(3915)}^{\gamma\gamma}\times\mathcal{B}[X(3915)\to \omega J/\psi]$ \cite{Zyla:2020zbs}
\begin{equation}
\begin{split}
\Gamma^{\gamma\gamma}_{X(3915)}\times\mathcal{B}[X(3915)\to \omega J/\psi]=54\pm9~{\rm eV},\label{aa}
\end{split}
\end{equation}
where the decay width of $\chi_{c0}(2P)\to \gamma\gamma$ had already been provided by different theoretical groups, which  are collected in Table~\ref{chi2doublegammafusion}. Thus,
the decay width of $\chi_{c0}(2P)\to \gamma\gamma$ channel can be averaged as $\Gamma^{\gamma\gamma}_{\chi_{c0}(2P)}=1.73\pm0.30$ keV, which can be used as input to estimate the following branching ratio
\begin{equation}\label{4eqBrchiomepsi}
\begin{split}
\mathcal{B}[\chi_{c0}(2P)\to \omega J/\psi]=(3.13\pm0.75)\times10^{-2}
\end{split}
\end{equation}
by combining with Eq. (\ref{aa}). Finally, we obtain
the branching fraction product for $\chi_{c0}(2P)$
\begin{equation}\label{4eqPBr}
\begin{split}
&\mathcal{B}[B\to K\chi_{c0}(2P)]\times\mathcal{B}[\chi_{c0}(2P)\to \omega J/\psi]\\
&=(1.1\pm0.6)\times10^{-5}.\\
\end{split}
\end{equation}

\begin{table}[htbp]
\caption{The calculated $\Gamma^{\gamma\gamma}_{\chi_{c0}(2P)}$ by different theoretical groups.}
\label{chi2doublegammafusion}
\renewcommand\arraystretch{1.6}
\begin{tabular*}{86mm}{@{\extracolsep{\fill}}ccccccc}
\toprule[1.0pt]
\toprule[1.0pt]
Reference&\cite{Li:2009zu}&\cite{Devlani:2013mky}&\cite{Bhatnagar:2018zzf}&\cite{Ebert:2003mu}&\cite{Munz:1996hb}&\cite{Wang:2007nb} \\
\toprule[0.8pt]
$\Gamma^{\gamma\gamma}_{\chi_{c0}(2P)}$~(keV)&1.7 &0.98 &1.91 &1.9 &1.110$\pm$0.130 &3.51\\
\bottomrule[1pt]
\bottomrule[1pt]
\end{tabular*}
\end{table}

In the following, we should mention another charmoniumlike $XYZ$ state $Y(3940)$, which was observed by the Belle Collaboration in the $B \to K \omega J/\psi$ process \cite{Abe:2004zs}. After three years, the BaBar Collaboration confirmed the $Y(3940)$ observation \cite{Aubert:2007vj}. Later, BaBar updated the measurement of the mass and the width of the $Y(3940)$, i.e., $m=3919.1^{+3.8}_{-3.5}({\rm stat})\pm2.0({\rm syst})$ MeV and $\Gamma=31^{+10}_{-8}({\rm stat})\pm5({\rm syst})$ MeV \cite{delAmoSanchez:2010jr}.  Since the resonant parameter of the $Y(3940)$ is close to that of the $X(3915)$, PDG  \cite{Nakamura:2010zzi} treated the $X(3915)$ and the $Y(3940)$ as the same state. In fact, this simple treatment should be proved by further theoretical investigation, which is a topic in this work.

When obtaining $\mathcal{B}[B\to K\chi_{c0}(2P)]\times\mathcal{B}[\chi_{c0}(2P)\to \omega J/\psi]$ shown in Eq.~(\ref{4eqPBr}), one may compare this result with the experimental data of $\mathcal{B}(B\to KY(3940))\times \mathcal{B}(Y(3940)\to J/\psi \omega)$. BaBar obtained
the branching fraction product $\mathcal{B}[B\to KY(3940)]\times \mathcal{B}[Y(3940)\to J/\psi \omega]=(7.1\pm1.3(\rm stat)\pm3.1(\rm syst))\times10^{-5}$~\cite{Abe:2004zs}. With more experimental data accumulated in the subsequent experiments, BaBar measured this product branching ratio again, which is
$\mathcal{B}[B\to KY(3940)]\times \mathcal{B}[Y(3940)\to J/\psi \omega]=(4.9^{+1.0}_{-0.9}(\rm stat)\pm0.5(\rm syst))\times10^{-5}$~\cite{Aubert:2007vj}.
The result of the product branching ratio for the $Y(3940)$ employed in the PDG is~\cite{delAmoSanchez:2010jr}
\begin{equation}\label{4eqPBr3940}
\begin{split}
&\mathcal{B}[B\to KY(3940)]\times \mathcal{B}[Y(3940)\to J/\psi \omega]\\
&=(3.0^{+0.7}_{-0.6}(\rm stat)^{+0.5}_{-0.3}(\rm syst))\times10^{-5}.\\
\end{split}
\end{equation}
If considering the experimental error, then we find that the obtained
$\mathcal{B}[B\to K\chi_{c0}(2P)]\times\mathcal{B}[\chi_{c0}(2P)\to \omega J/\psi]$ in this work is comparable with the experimental data of $\mathcal{B}[B\to KY(3940)]\times \mathcal{B}[Y(3940)\to J/\psi \omega]$. Thus, it is possible to assign the $Y(3940)$ to be the $\chi_{c0}(2P)$ state. To draw definite conclusion on this point we have to wait for more precise experimental data of $B\to K J/\psi\omega$ in near future. Here, LHCb and Belle II  will have good opportunity.

Having the above discussion, we should close the present work with a short summary. In this work, we extract the branching ratio $\mathcal{B}[B\to K\chi_{c0}(2P)]$ for the first time through the fit fractions in the newly measured $B\to KD\bar{D}$ process by LHCb \cite{Aaij:2020ypa, Aaij:2020hon}. Focusing on this obtained branching ratio, we further perform a calculation of $\mathcal{B}[B\to K\chi_{c0}(2P)]$ by the rescattering mechanism, and find that this extracted branching ratio is reproduced well. By this study, indeed  the newly measured $B\to KD\bar{D}$ process plays a crucial role to establish the $\chi_{c0}(2P)$ charmonium state.

\section*{Acknowledgements}
This work is supported by the China National Funds for Distinguished Young Scientists under Grant No. 11825503, the National Key Research and Development Program of China under Contract No. 2020YFA0406400, the 111 Project under Grant No. B20063, the National Natural Science Foundation of China under Grans No. 12047501, and the Fundamental Research Funds for the Central Universities.


\end{document}